\documentstyle[12pt]{article}
\input boxedeps.tex
\SetRokickiEPSFSpecial  
\HideDisplacementBoxes
\begin{document}
\mbox{}\hfill MPI H-V38-1995\\
\mbox{}\hfill hep-th/9510104\\

\begin{center}
{\Large Tube model for light-front QCD}
\vspace{.1in}

Brett van de Sande\\
Max Planck Institut f\"ur Kernphysik,\\
Postfach 10.39.80, D-69029 Heidelberg, Germany
\vspace{.1in}

and
\vspace{.1in}

Matthias Burkardt\\
Physics Department\\
New Mexico State University\\
Las Cruces, NM 88003-0001, U.S.A.
\vspace{.1in}
\end{center}

\begin{abstract}
We propose the tube model as a first step in solving the
bound state problem in light-front QCD.  In this approach
we neglect transverse variations of the fields, producing
a model with 1+1 dimensional dynamics.  We then solve the
two, three, and four particle sectors of the
model for the case of pure glue SU(3).
We study convergence to the continuum limit and various
properties of the spectrum.
\end{abstract}

The ultimate goal of light-front field theory is to start with the
QCD Lagrangian and, with a minimum of approximation, calculate the
hadron spectrum.  The basic idea behind this approach is to use
Hamiltonian techniques in the
coordinate system best suited for relativistic dynamics~\cite{pw}.
We believe that the  ``tube model'' provides a good
starting point towards that goal.
Simply put, we take full 3+1 dimensional QCD and restrict our
attention to a region that is small in the two transverse
directions and is large in the longitudinal dimension,
hence the name ``tube model.''
Since the transverse region is small, we neglect any transverse
derivatives of the fields as an approximation.
The resulting theory is a 1+1 dimensional
gauge theory coupled to adjoint matter.
We then employ light-front Hamiltonian techniques to
produce the low energy spectrum of the theory.
This approach was
first explored in the transverse lattice paper of
Bardeen, Pearson, and Rabinovici~\cite{bardeen}
(that is, one can think of the tube model as two links
on a transverse lattice) and has become
increasingly popular in the last few years
under the name ``dimensional reduction''~\cite{adjoint,dalley}.

The tube model has the virtues of a 1+1 dimensional theory,
in that it is superrenormalizable and computationally tractable,
while retaining some of the dynamics of the full theory.
Thus, as we demonstrate in this paper, one can hope
to understand the behavior of one tube quite thoroughly.
The subsequent step will be much more difficult:  to
couple several of these tubes together to make a full
3+1 dimensional theory.

In this paper, we study only the pure glue sector of the
theory.  Of course, an important next step will be to
include fermions.  Although one may impose antiperiodic
boundary conditions on the fermion fields, a consistent formulation
that includes fermions must impose periodic boundary conditions
on the gauge fields.
In anticipation of this situation, we impose periodic boundary
conditions for our gauge fields;
this will have some important consequences for our numerical results.

The paper is organized as follows.  In Sec.\ \ref{gzm} we
discuss (and dismiss) the gauge zero mode.  We define the
tube model for pure glue in Sec.\ \ref{pg}.  Then in Secs.\ \ref{twop},
\ref{threep}, and \ref{fourp} we discuss the two, three, and
four particle sectors of the theory.
In addition, our technique for
measuring parity is demonstrated in Sec.\ \ref{threep}.
Finally, we discuss some aspects of
our results in Sec.\ \ref{discussion}.

\section{The gauge zero mode}
\label{gzm}

Before we begin, we must discuss the ``gauge zero mode,''
$V \equiv \int dx^-\, A_- $.
As has been noticed by several authors in the context of both
light-front~\cite{many} and equal time~\cite{multitude} quantization,
one cannot set $V=0$ by choice of gauge because it is a genuine
dynamical degree of freedom.  Only the gauge choice
$\partial_- A_- =0$ is allowed.

What are the effects of $V$ on the tube model particle spectrum?
In the Lagrangian, we must replace the longitudinal derivative
$\partial_-$ by the covariant version $\partial_- + i g V$.
If we regulate the theory by imposing periodic boundary
conditions (DLCQ), the longitudinal momentum becomes discrete
and we find that the effect of $V$ is to shift momenta by an
amount that is less than one longitudinal momentum spacing.
Clearly, any direct effect of this mode on ordinary dynamical
modes is negligible in the large volume limit $L \to \infty$.
However, the zero mode sector of the theory may become qualitatively
different.

For instance, some of the zero
modes that were otherwise constrained become dynamical.
In light-front field theory, equations of motion
are typically of the form
$\partial_{+} (\partial_- + i g V) \phi = U(\phi)$,
where $U(\phi)$ is the potential term.  When $V=0$ it is clear why zero
modes $\partial_- \phi=0$ are not dynamical:  they have no conjugate momentum.
However, note that when $V$ is nonzero, the associated kinetic
term is small (since $V$ is small) relative to the kinetic term of any
non-zero mode.
Thus, one may just as well treat these zero
modes as non-dynamical (that is, as a solution of a constraint equation).

Without the gauge zero mode, Gau{\ss}' law
requires that all $N^2-1$ generators $Q^a$ of the gauge group
(defined in Eqn.~(\ref{charge})) annihilate physical states.  If $V$ is
included, only the $N-1$ diagonal generators
must annihilate physical states.  Consequently, the
spectrum contains many more
``color singlet'' states if $V$ is nonzero.  However, one finds
that these additional states have eigenvalues that diverge as one
takes the continuum limit $L\to \infty$ and they do not affect the low
energy spectrum \cite{rolf,schreiber}.

Although $V$ may be quite important for understanding the vacuum
state itself, its effects on the particle spectrum seem to be
small in the type of model we are studying.
Thus, we set $V=0$ as a dynamical approximation.  It is not clear
that $A_-$ can be so easily neglected in the full 3+1 dimensional theory.
For instance, in 3+1 dimensions, $A_-$ has a ``proper zero mode,''
$\int dx^-\, A_-$,
that is constrained along with a dynamical ``global zero mode,''
$\int dx^- \, d^2 x^r A_-$.
The effects of the proper zero mode on the particle spectrum could be
quite important~\cite{alex}.

\section{The Hamiltonian}
\label{pg}
In this section, we define the tube model for 3+1 dimensional
$SU(N)$ gauge theory, introduce our notation, and write
out the light-front Hamiltonian.
We introduce light front
coordinates $x_\mp=x^\pm = (x^0 \pm x^3)/\sqrt{2}$,
where $x^+$ is light-front ``time'' and $x^-$ is the longitudinal
spatial coordinate; transverse indices are labeled by $r,s \in
\left\{1,2\right\}$.  Next, we put the system in a box of dimensions
$(L,L_{\perp},L_{\perp})$ in the $(x^-, x^1, x^2)$ directions and impose
periodic boundary conditions, or ``discretized light-cone
quantization'' (DLCQ) \cite{pw,pb}.
As our central assumption, we neglect variations
of the fields in the transverse directions and set  $\partial_r A_\mu = 0$.
Since this approximation violates transverse gauge invariance and rotational
invariance, we are free to introduce
a transverse mass term ${\rm tr} A_r A_r$.
As we shall see, inclusion of this term
will be necessary for proper
renormalization of the model.  Thus the Lagrangian is
\begin{equation}
L_{\rm gauge} = L_{\perp}^2 \int dx^- \, {\rm tr}\left\{
    F_{-+} F_{-+}+2 F_{+r}F_{-r} -\frac{1}{2} F_{rs}F_{rs} -
    \mu_0^2 A_r A_r
  \right\} \; ,
\end{equation}
where
\begin{eqnarray}
     F_{-+}&=& \partial_-A_+ \\
     F_{-r} &=& \partial_- A_r \\
     F_{+r} &=& \partial_+ A_r+ i g\left[A_+,A_r\right] \\
     F_{rs}&=& i g \left[A_r,A_s\right] \; .
\end{eqnarray}

The $A_+$ field is a constrained degree of freedom.  It obeys the
constraint equation
\begin{equation}
 - \frac{1}{g} \partial_- \partial_- A_+ = J^+ = i\left[
                \partial_- A_r, A_r \right] \; ,
		\label{constraintplus}
\end{equation}
where $J^\mu$ is the conserved current.  The conserved charge associated
with $J^\mu$ is
\begin{equation}
       Q = L_{\perp}^2 \int dx^- \, J^+ =
       i L_{\perp}^2 \int dx^- \left[
                       \partial_- A_r, A_r \right] \; . \label{charge}
\end{equation}
According to Eqn.~(\ref{constraintplus}), $Q$ must vanish.  In the usual
manner, we demand that $Q$ annihilates physical states.

The generators of $SU(N)$ obey the relations
$\left[T^a,T^b\right] = i f^{a b c} T^c$ and
${\rm tr}\, T^a T^b = \delta^{a b}/2$.
Writing $A_r = A_r^a T^a$,
the momentum conjugate to $A_r^a$ is
$\partial_- A_r^a$.
Thus the zero mode of $A_r^a$ has no conjugate
momentum and is not a true dynamical degree of freedom.
For the rest of this discussion, we will drop the zero mode
associated with $A_r$.
Next, we construct the
helicity operator ${\cal J}_z$ which generates rotations in the $(x^1,x^2)$
plane,
the longitudinal momentum operator $P^+$, and the Hamiltonian $P^-$:
\begin{eqnarray}
  {\cal J}_z &=&
	-4 L_{\perp}^2 \int dx^- {\rm tr}\, A_1 \, \partial_-A_2 \; ,
\\
P^+ &=&
	2 L_{\perp}^2 \int dx^- {\rm tr}\, \partial_- A_r \, \partial_- A_r
 \; ,\\
P^- &=&
	L_{\perp}^2 \int dx^- {\rm tr} \left\{
	-g^2 J^+ \frac{1}{\left(\partial_-\right)^2} J^+ +
	\mu_0^2 A_r A_r -\frac{g^2}{2}{\left[A_r,A_s\right]}^2
	\right\} \; . \label{ham}
\end{eqnarray}
The inverse derivative operator is well-defined in this case since
$J^+$ has no zero mode in the space of physical states.

Now we are ready to quantize the theory.  We can expand $A_r$ in
terms of creation and annihilation operators:
\begin{equation}
A_r^a = \frac{1}{L_{\perp} \sqrt{4 \pi}}
	\sum_{\lambda=\pm 1}
        \sum_{k=1}^\Lambda \frac{1}{\sqrt{k}}\left(
	a_{k,\lambda}^a \epsilon_r^{-\lambda} e^{-2 \pi i k x^-/L}
      +{a_{k,\lambda}^a}^{\dag} \epsilon_r^{\lambda}
	                e^{2 \pi i k x^-/L}
	\right) \; .
\end{equation}
The creation/annihilation operators obey the usual commutation
relations,
\begin{eqnarray}
 \left[a_{k,\lambda}^a,{a_{l,\lambda^\prime}^b}^{\dag}\right]
  &=& \delta^{a,b} \delta_{k,l}\delta_{\lambda,\lambda^\prime} \\
 \left[a_{k,\lambda}^a,a_{l,\lambda^\prime}^b\right]
 &=&\left[{a_{k,\lambda}^a}^{\dag},{a_{l,\lambda^\prime}^b}^{\dag}\right]
  = 0 \; .
\end{eqnarray}
We define the polarization tensor $\epsilon_r^{\lambda}$:
$\epsilon_1^{\pm 1}=1/\sqrt{2}$ and
$\epsilon_2^{\pm 1}=\pm i/\sqrt{2}$.
Using the creation/annihilation operators and normal ordering, we find
\begin{eqnarray}
    {\cal J}_z = \sum_{k=1}^\Lambda \lambda \,
	{a_{k,\lambda}^a}^{\dag} a_{k,\lambda}^a \\
P^+ = \frac{2 \pi}{L} \sum_{k}  k\,
        {a_{k,\lambda}^a}^{\dag} a_{k,\lambda}^a \\
Q^a = -i \sum_{k=1}^\Lambda f^{a b c}
        {a_{k,\lambda}^b}^{\dag} a_{k,\lambda}^c
             \; .
\end{eqnarray}

Let us examine the
Hamiltonian $P^-$ in more detail.  From the first two terms
in Eqn.~(\ref{ham}) we obtain the four instantaneous interactions,
the ``self-induced inertias,'' and the mass term,
\begin{eqnarray}
P^-_1 &=&
{{g^2 L}\over {16 {\pi }^2 L_{\perp}^2 }}
    \sum_{n^\prime,m^\prime,n,m}
  {{{\sqrt{n n^\prime}}\delta_{m+n,m^\prime + n^\prime}
      f^{a b c}f^{a^\prime b^\prime c}
      }\over
    {{\sqrt{m m^\prime}}{{\left( m + n \right) }^2}}}
    \nonumber \\ & & \hspace{1.5in} \cdot
    	{a_{n,-\lambda}^{a}}^{\dag}
       {a_{m,\lambda}^{b}}^{\dag}
        a_{n^\prime,\lambda^\prime}^{a^\prime}
         a_{m^\prime,-\lambda^\prime}^{b^\prime}
	\label{h1}
\\
P^-_2 &=&
{{g^2 L}\over {16 {\pi }^2 L_{\perp}^2 }}
    \sum_{n^\prime,m^\prime,n,m}
  {{\left( m+n \right)
      {\sqrt{n^\prime}}\delta_{n,m + m^\prime + n^\prime}f^{a b c}
      f^{a^\prime b^\prime c}}\over
    {{\sqrt{m n m^\prime}}{{\left( m^\prime + n^\prime \right) }^2}}}
    \nonumber \\ & & \hspace{1in} \cdot
    {a_{n,\lambda}^{a}}^{\dag}
       a_{n^\prime,\lambda^\prime}^{a^\prime}
        a_{m^\prime,-\lambda^\prime}^{b^\prime} a_{m,\lambda}^{b}
\\
P^-_3 & =& -
{{g^2 L }\over {16 {\pi }^2 L_{\perp}^2 }}
     \sum_{n^\prime,m^\prime,n,m}
  {{{\sqrt{n}}\left( m^\prime + n^\prime \right)
      \delta_{m+n+n^\prime,m^\prime}}
     \over {{\sqrt{m^\prime n^\prime m}}{{\left( m + n \right) }^2}}}
    \nonumber \\ & & \hspace{1.5in} \cdot
    f^{a b c}
      f^{a^\prime b^\prime c}
     {a_{n^\prime,\lambda^\prime}^{a^\prime}}^{\dag}
        {a_{n,-\lambda}^{a}}^{\dag}
        {a_{m,\lambda}^{b}}^{\dag} a_{m^\prime,\lambda^\prime}^{b^\prime}
\\
P^-_4 &=& -
{{ g^2 L }\over {16 \pi^2  L_{\perp}^2}}
   \sum_{n^\prime,m^\prime,n,m,k>0}
  {{\left(m+n \right) \left(m^\prime +n^\prime \right) }\over
     {\sqrt{m n m^\prime n^\prime}{k^2}}}
        \nonumber \\ & & \hspace{0.5in} \cdot
	\delta_{n,k + m}
      \delta_{m + m^\prime , n + n^\prime}
	f^{a b c}
      f^{a^\prime b^\prime c}
    {a_{n^\prime,\lambda^\prime}^{a^\prime}}^{\dag}
        {a_{n,\lambda}^{a}}^{\dag}
        a_{m^\prime,\lambda^\prime}^{b^\prime} a_{m,\lambda}^{b}
	\label{h4}
\\
P^-_5 &=&
{{ g^2 L N}\over {32 {\pi }^2  L_{\perp}^2}}
     \sum_{m}
    f\!\left(m\right)
   {a_{m,\lambda}^{b}}^{\dag} a_{m,\lambda}^{b}
\\
P^-_6 &=&
{{L \mu^2}\over {4\pi }}
     \sum_{k}
    {{{a_{k,\lambda}^{c}}^{\dag} a_{k,\lambda}^{c}}\over k}
  \; . \label{mass}
\end{eqnarray}
The self-induced inertia $P^-_5$ is obtained by bringing the
instantaneous interactions in Eqn.~(\ref{ham})
into normal ordered form, Eqns.~(\ref{h1}) and (\ref{h4}).
Before mass renormalization, the associated function $f(m)$ is
\begin{equation}
f(m)= 8 \sum_{k=1}^{m-1} \frac{1}{k^2}+
	\frac{2}{m}\left(\ln(\Lambda)+\gamma_E\right)+
	\frac{3}{m^2}+O\left(1/\Lambda\right) \; .
\end{equation}
We choose our mass renormalization such that the logarithmic
divergence is removed
\begin{equation}
f(m) = 8 \sum_{k=1}^{m-1} \frac{1}{k^2}
	+\frac{3}{m^2}+\frac{8}{m} \approx \frac{4 \pi^2}{3} \; .
\end{equation}
Next, we can write out terms associated with
$\left[A_r,A_s\right]^2$.  Any contracted
terms have the same operator structure as the mass term
and are removed by mass renormalization.
\begin{eqnarray}
P^-_7 &=&
{{{g^2}L}\over {16{{\pi }^2  L_{\perp}^2 }}}
     \sum_{k,l,m}
  {{f^{a b c}f^{a^\prime b^\prime c}
      }\over
    {{\sqrt{k l m \left(k + l + m\right)}}}}
    \nonumber \\ & & \hspace{1.5in} \cdot
    {a_{k + l + m,\lambda^\prime}^{a}}^{\dag}
       a_{k,\lambda}^{b} a_{l,\lambda^\prime}^{a^\prime}
         a_{m,-\lambda}^{b^\prime}
\\
P^-_8 &=&
{{{g^2}L}\over {16{{\pi }^2 L_{\perp}^2 }}}
    \sum_{k,l,m}
  {{f^{a b c}f^{a^\prime b^\prime c}
      }\over
    {{\sqrt{k l m \left(k + l + m\right)}}}}
    \nonumber \\ & & \hspace{1.5in} \cdot
    {a_{m,-\lambda}^{b^\prime}}^{\dag}
       {a_{l,\lambda^\prime}^{a^\prime}}^{\dag}
        {a_{k,\lambda}^{b}}^{\dag} a_{k + l + m,\lambda^\prime}^{a}
\\
P^-_9 &=&
{{{g^2}L}\over {32{{\pi }^2 L_{\perp}^2 }}}
    \sum_{k,l,m,n}
  {{\delta_{m+n,k + l}f^{a b c}f^{a^\prime b^\prime c}
      }\over
    {{\sqrt{k l m n}}}}
    \nonumber \\ & & \hspace{1.5in} \cdot
    {a_{k,\lambda^\prime}^{a}}^{\dag}
       {a_{l,\lambda}^{b}}^{\dag}
        a_{m,\lambda^\prime}^{a^\prime} a_{n,\lambda}^{b^\prime}
\\
P^-_{10} &=&
{{{g^2}L}\over {32{{\pi }^2 L_{\perp}^2 }}}
     \sum_{k,l,m,n}
  {{\delta_{l+n,k +m}f^{a b c}f^{a^\prime b^\prime c}
      }\over
    {{\sqrt{k l m n}}}}
    \nonumber \\ & & \hspace{1.5in} \cdot
    {a_{k,{\lambda^\prime}}^{a}}^{\dag}
       {a_{m,-{\lambda^\prime}}^{a^\prime}}^{\dag}
        a_{l,\lambda}^{b} a_{n,-\lambda}^{b^\prime}
\\
P^-_{11} &=&
{{{g^2}L}\over {32{{\pi }^2 L_{\perp}^2 }}}
     \sum_{k,l,m,n}
  {{\delta_{l+m,k + n}f^{a b c}f^{a^\prime b^\prime c}
      }\over
    {{\sqrt{k l m n}}}}
    \nonumber \\ & & \hspace{1.5in} \cdot
    {a_{k,\lambda}^{a}}^{\dag}
       {a_{n,\lambda^\prime}^{b^\prime}}^{\dag}
        a_{l,\lambda^\prime}^{b} a_{m,\lambda}^{a^\prime}
\; .
\end{eqnarray}
The relation between the bare mass $\mu_0$ and the renormalized
mass $\mu$ in Eqn.~(\ref{mass}) is
\begin{equation}
 \mu_0^2 = \mu^2 + \frac{g^2 N}{4 \pi L_{\perp}^2} -
            \frac{ g^2 N}{2 \pi L_{\perp}^2}
            \sum_{m=1}^\Lambda \frac{1}{m}
	    \; .\label{renorm}
\end{equation}

In the subsequent numerical work, we choose
\begin{equation}
 1= \frac{g^2 N}{4 \pi L_{\perp}^2}
\end{equation}
which sets the mass scale.  For the three and four particle truncations
where we need to specify the number of colors we set $N=3$.
Also, we define $M^2$ as the
eigenvalue of the $(\mbox{invariant mass})^2$ operator
$P_\mu P^\mu = 2 P^+ P^-$.  If we define the integer valued
``harmonic resolution'' $K=L P^+/(2 \pi)$, the continuum
limit is achieved by taking $\lim K \to \infty$ for fixed $P^+$.
States are classified by the conserved quantities:
helicity ${\cal J}_z$, particle number mod 2,
charge conjugation ${\cal C}$, and parity ${\cal P}$.
We will often use the notation ${{\cal J}_z}^{\cal PC}$.
Under charge conjugation, gauge fields transform as
$\left(A_r\right)_{ij} \to -\left(A_r\right)_{ji}$
and under parity, they transform as $A_r(x^+,x^-) \to -A_r(x^-,x^+)$
(actually, $P^+$ eigenstates are eigenstates of parity times a
longitudinal boost).
Note that we use the definition of parity appropriate for 3+1 dimensions.
Our method for measuring parity will
be explained in more detail in Section~\ref{threep}.

\section{Two particles}
\label{twop}

The low energy states in this type of model have,
to a good approximation, definite particle number~\cite{adjoint}.
Thus, it is useful to study the spectrum by applying a truncation
in particle number.
%
%
\begin{figure}
\centering
\BoxedEPSF{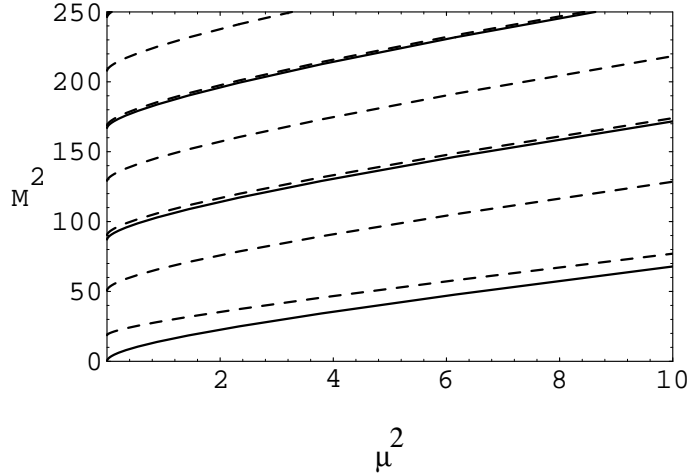 scaled 1200}
\caption{Two particle spectrum vs $\mu^2$ for ${\cal J}_z=0$
(dashed curves) and
  ${\cal J}_z=2$ (solid curves) using the polynomial  wave function
  basis calculation discussed at the end of Sec.\ \protect\ref{twop}.
  For ${\cal J}_z=0$, the states alternate between even
  and odd parity.  \label{spec3}}
\end{figure}
In this section, we apply a two particle truncation
and study the bound state equation using
DLCQ and a polynomial wave function basis in the continuum limit.

\begin{figure}
\centering
\BoxedEPSF{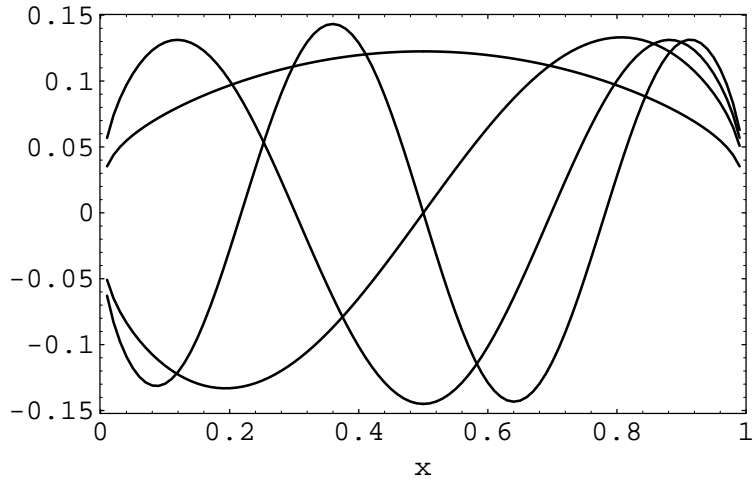 scaled 1200}
\caption{Lowest eigenfunctions $\phi_j$
vs momentum fraction $x=j/K$ for ${\cal J}_z=0$, $\mu^2=0$,
and $K=100$. Wavefunctions are either even or odd, depending
on the parity of the state. \label{wave0}}
\end{figure}

\begin{figure}
\centering
\BoxedEPSF{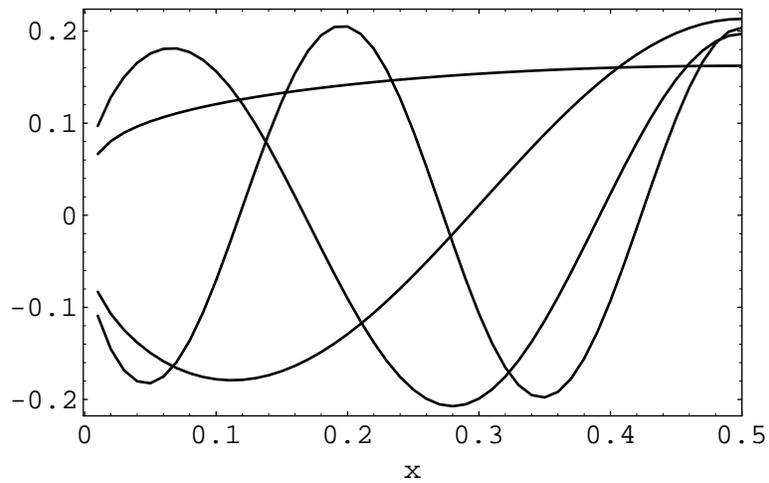 scaled 1200}
\caption{Lowest eigenfunctions $\phi_j$
vs momentum fraction $x=j/K$ for ${\cal J}_z=2$, $\mu^2=0$,
and $K=100$.  The shape of the lowest wavefunction is a
finite $K$ artifact; it becomes constant in the continuum
limit $K \to \infty$. \label{wave2}}
\end{figure}

Two particle eigenstates have the general form
\begin{equation}
        \sum_{j=1}^{K-1} \phi_j \, {a_{j,\lambda}^a}^{\dag}
               \, {a_{K-j,\lambda^\prime}^a}^{\dag}\, | 0 \rangle
\end{equation}
where the total helicity is ${\cal J}_z= \lambda + \lambda^\prime $.
These states are even under charge conjugation ${\cal C}= 1$.
For ${\cal J}_z = 2$, the wavefunction must be even,
$\phi_j = \phi_{K-j}$, since the particles are identical
(${{\cal J}_z}^{\cal PC}=2^{++}$), while they can be even or odd
for ${\cal J}_z = 0$ (${{\cal J}_z}^{\cal PC} = 0^{++}$ or $0^{-+}$).
The spectrum as a function of mass is shown in Fig.~\ref{spec3}.
The states are roughly evenly spaced
due to the linear potential generated by the instantaneous interactions
$P_1^- + \cdots + P_4^-$.
In addition, for $\mu^2=0$, it is evident that the ground
state is massless in the continuum limit.
The corresponding wavefunctions at $\mu^2=0$ are roughly sines
and cosines as shown in Figs.~\ref{wave0} and \ref{wave2}.
%
%
%
%

Taking the continuum limit $K\to\infty$, one can convert
the bound state equation into an integral equation
\begin{eqnarray}
  M^2 \phi(x) &=& \mu^2 \phi(x) \left( \frac{1}{x}+\frac{1}{1-x}\right)
  \nonumber \\ & &
         - \frac{g^2 N}{4 \pi L_{\perp}^2} \int
	 \frac{ dy \, \left(x+y\right)\left(2-x-y\right)}
	 {(x-y)^2 \sqrt{x (1-x) y (1-y)}} \phi(y) \nonumber \\
   &  & \mp \frac{g^2 N}{4 \pi L_{\perp}^2}\int
            \frac{ dy }{\sqrt{x (1-x) y (1-y)}} \phi(y) + O(1/K) \; .
  \label{bse}
\end{eqnarray}
where $x,y \in \left(0,1\right)$ and we identify $\phi(i/K)$ with $\phi_i$
in the discrete case.
For ${\cal J}_z=2$, the coefficient of the last term
of (\ref{bse})
is negative and $\phi(x)=\phi(1-x)$. For ${\cal J}_z=0$,
the coefficient of the last term is positive and $\phi(x)=\pm \phi(1-x)$.
The renormalization prescription we introduced for the discrete case
Eqn.~(\ref{renorm})
is equivalent to using a principal-value prescription
for the pole in the first integral of Eqn.~(\ref{bse}).
One finds that $\phi(x)=1$, $M^2=0$ is an exact solution for
${\cal J}_z=2$ and $\mu^2=0$, confirming our numerical
results.  Below this mass, the spectrum
becomes tachyonic independent of the $\left[A_r,A_s\right]^2$
interaction.  Also, we can repeat 't Hooft's end-point analysis
\cite{bardeen,thooft} for this state.
Assume that the wavefunction has power law
behavior $\phi(x) \propto x^\beta$ near the endpoint $x=0$.
To leading order in $x$, the bound state equation becomes
\begin{equation}
  0= \mu^2 x^{\beta-1} - \frac{g^2 N}{2 \pi L_{\perp}^2} \int_0^1
  	\frac{dy \, (x+y) y^{\beta-1/2}}{(x-y)^2 \sqrt{x}}
	\; ,
\end{equation}
which yields the relation
\begin{equation}
  \mu^2 = \frac{g^2 N}{\pi L_{\perp}^2} \pi \beta
  	\tan \! \left(\pi \beta\right) \, , \;\;\;\;  0 \le \beta \le 1/2 \; .
     \label{endpoint}
\end{equation}
Using this expression for the endpoint behavior, one can also
show that $M^2$ depends linearly on $\mu$ near $\mu^2=0$:
\begin{equation}
\lim_{\mu^2 \to 0} M^2 = 4\pi \mu \sqrt{\frac{g^2N}{4\pi L_\perp^2}} \; .
\end{equation}

\begin{figure}
\centering
\BoxedEPSF{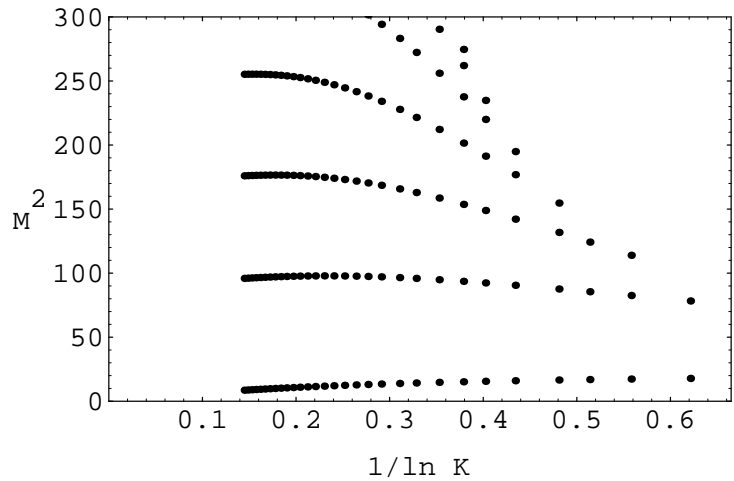 scaled 1200}
\caption{Two particle spectrum for helicity 2, $\mu^2=0$, as
a function of harmonic resolution.  According to this plot, convergence
to the continuum limit seems to be quite good. \label{spec1}}
\end{figure}

Next, let us examine the convergence to the continuum limit
for DLCQ.  If we plot the spectrum as a function of
harmonic resolution, we seem to find quick convergence as
shown in Fig.~\ref{spec1}.
%
%
%
%
%
\begin{figure}
\centering
\BoxedEPSF{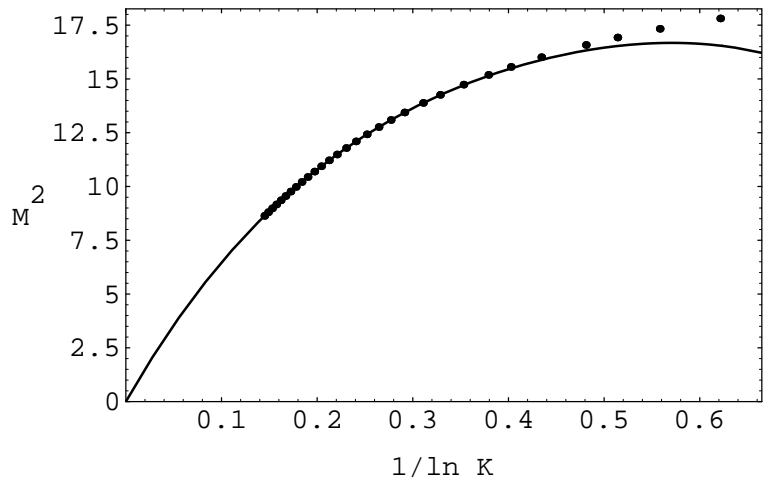 scaled 1200}
\caption{Ground state eigenvalue for helicity 2, $\mu^2=0$ as
a function of harmonic resolution together with a numerical
fit to $78.08 /\ln K - 150.52 /(\ln K)^2  + 180.21 /(\ln K)^3
+ \ldots$.  In fact, convergence to the continuum limit
is quite poor due to our having neglected the zero mode.
\label{spec2}}
\end{figure}
However, if we examine the ground state more closely in
Fig.~\ref{spec2}, we see
that its eigenvalue in fact converges quite slowly.
The same is true for the eigenfunctions.
Let us take a variational ansatz for
the wavefunction:  $\phi_j = j^\beta (K-j)^\beta$.
For ${\cal J}_z=2$, $\mu^2=0$, we find
that this ansatz is a very good approximation
of the ground state wavefunction for all $K$
($M^2$ eigenvalues differ by only
about 0.002) and that the wavefunction is constant ($\beta \to 0$)
in the $K \to \infty$ limit:
%
%
%
\begin{equation}
\beta \approx 1.71 /\ln K  - 2.95 /(\ln K)^2 + 5.49/(\ln K)^3  + \ldots
 \; .
\end{equation}
This explains the shape of the lowest eigenfunction in
Figs.~\ref{wave2} and \ref{wave3}.
%
%
\begin{figure}
\centering
\BoxedEPSF{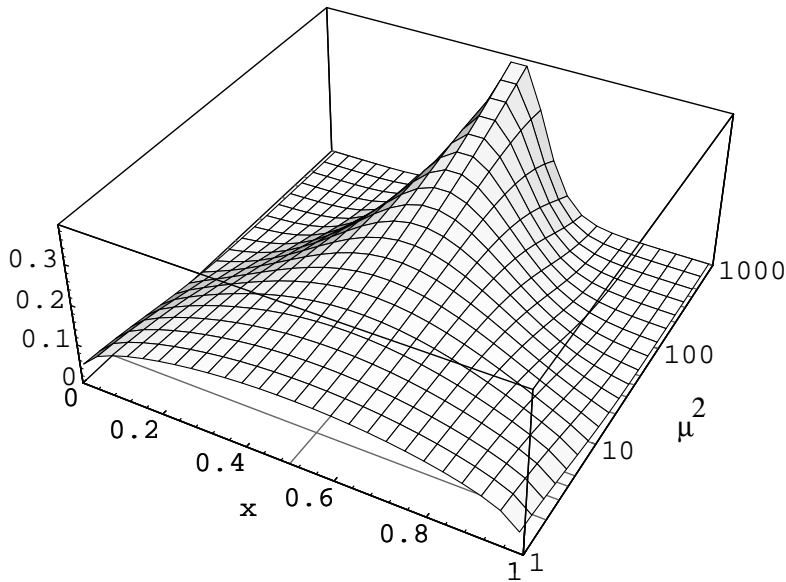 scaled 1200}
\caption{Lowest eigenfunction vs $\mu^2$ for ${\cal J}_z=2$
and $K=102$.  The wavefunction is constant
at $\mu^2=0$ and becomes a delta function at $\mu^2=\infty$.
\label{wave3}}
\end{figure}

This slow convergence is due to our having treated the zero
mode sector of the theory poorly:  we have simply removed the zero
momentum mode from the theory.  In terms of
the continuum formulation, Eqn.~(\ref{bse}),
neglecting the zero mode is equivalent to introducing cutoffs in
momenta.  That is,
\begin{eqnarray}
 \varepsilon \le x, y \le 1-\varepsilon \; ,  \nonumber \\
 \varepsilon \le \left|x-y\right| \; ,
\end{eqnarray}
where we identify $\varepsilon \approx 2/K$.  In this case
the pole associated
with the first integral is removed by adding a term of
the form
\begin{equation}
  \frac{2 g^2 N}{\pi L_{\perp}^2} \frac{\phi(x)}{\varepsilon}
\end{equation}
to the right hand side of (\ref{bse}).  Using the variational ansatz
$\phi(x) = x^\beta (1-x)^\beta$, for ${\cal J}_z=2$,
$\mu^2=0$, we find the same behavior that we did in the discrete
case:
%
%
%
%
\begin{eqnarray}
	\beta &\approx& -\frac{1.838}{\ln \varepsilon}
	 - \frac{3.399}{\left(\ln \varepsilon\right)^2}
	 - \frac{12.611}{\left(\ln \varepsilon\right)^3} + \ldots \\
	\langle \phi |P_\mu P^\mu |\phi\rangle &\approx&
          -\frac{85.15}{\ln\varepsilon}
	  - \frac{197.96}{\left(\ln \varepsilon\right)^2}
	  - \frac{518.98}{\left(\ln \varepsilon\right)^3} + \ldots \\
\end{eqnarray}
This slow convergence problem gradually disappears as
$\mu^2$ is increased since the amplitude of the wavefunctions
becomes small near $x=0$ and $1$; see Fig.~\ref{wave3}.

In light of the problems associated with DLCQ, we introduce
a polynomial wave function basis \cite{bardeen}
to calculate the spectrum using
the continuum limit of the bound state equation (\ref{bse}).
We assume the wavefunction is of the form
$\phi(x)= x^\beta (1-x)^\beta P_n(x)$, where $P_n(x)$ is a
polynomial of order $n$ and $\beta$ is given
by Eqn.~(\ref{endpoint}).  This method is extremely accurate.
For example, with only five polynomials in the
basis we obtain a relative error of between
$10^{-4}$ ($\mu^2 \approx 0$) and
$10^{-3}$ ($\mu^2 \approx 10$) for the ground state eigenvalue.

\section{Three particles}
\label{threep}

For the three and four particle truncations, we use only the DLCQ approach
since it would be difficult to implement a wave function basis.
Unfortunately, the slow convergence problem that we saw in the two
particle truncation will affect the three and four particle spectra
as well.  In fact, the results we present for finite $K$ and $\mu^2=0$
better reflect the behavior of the model for small but nonzero
$\mu^2$ in the continuum limit.  That is, the dominant
finite $K$ effects can be absorbed into a finite mass renormalization.
In any case, the numbers we present should not be considered to be precise.

%
\begin{figure}
\centering
\BoxedEPSF{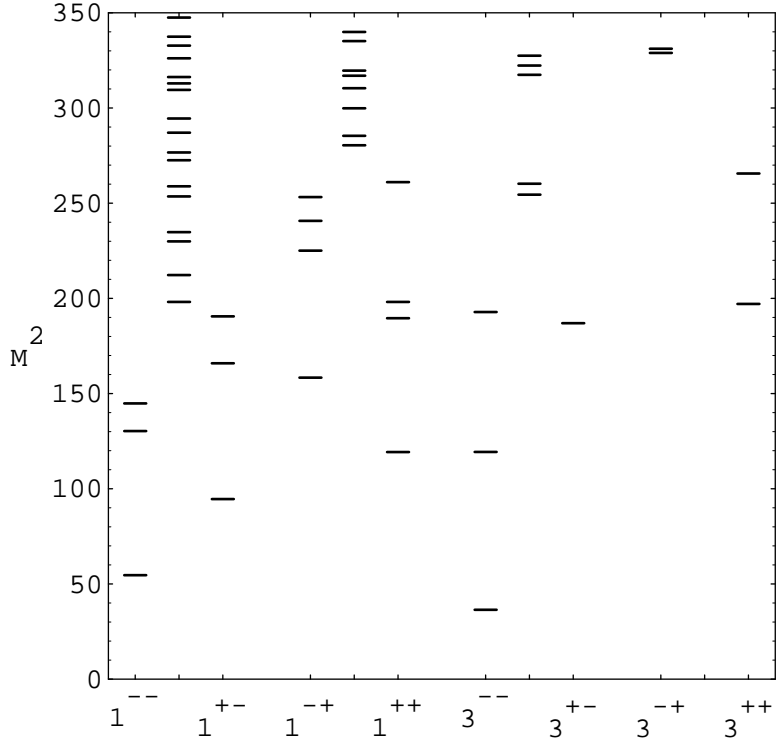 scaled 1200}
\caption{Three particle spectrum for various values of
${{\cal J}_z}^{\cal P C}$ where $\mu^2=0$ and $K=25$ (${\cal J}_z=1$)
or $K=35$ (${\cal J}_z=3$).  States whose parity could not be determined
are placed midway between  the corresponding
${{\cal J}_z}^{\cal +C}$ and ${{\cal J}_z}^{\cal -C}$ columns.
\label{threespec}}
\end{figure}
The three particle spectrum is shown in
Fig.~\ref{threespec}.
Color singlet states are constructed from the symmetric tensor
$d_{a b c}$ (${\cal C}=-1$) and the antisymmetric tensor
$f_{a b c} $ (${\cal C}=1$) \cite{color}.
The spectrum behaves much in the same way that atomic spectra do:  the
greater the symmetry of the wavefunction, the lower the energy of
the state.  Thus, the ${\cal J}_z=3$ sector has lower energy than
the ${\cal J}_z = 1$ sector;
${\cal C}=-1$ ($d_{abc}$) states have lower energy than
${\cal C}=1$ ($f_{abc}$) states;
and ${\cal PC}=1$ states have lower energy than ${\cal PC}=-1$
states.  In fact, we find the same basic behavior in the two particle
sector and, for the most part,
in the four particle sector as well.\\

%
\begin{figure}
\centering
\BoxedEPSF{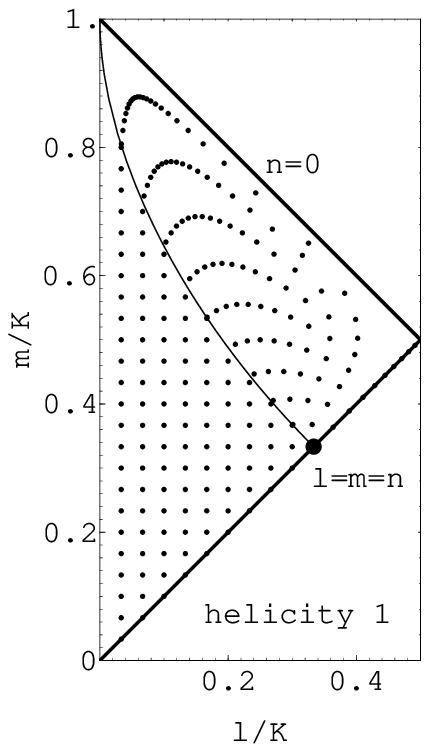 scaled 1200} \BoxedEPSF{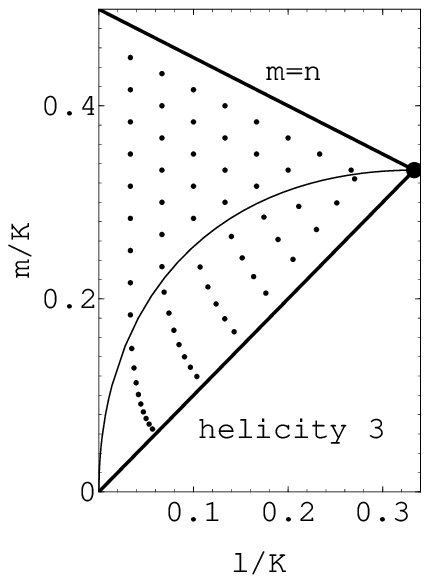 scaled 1200}
\caption{Hornbostel's parity transform applied to
three particle wavefunctions in Eqns.\
(\protect\ref{b1}) and (\protect\ref{b3}).
The transform maps each momentum combination $(l,m)$ into
another momentum combination $(\tilde{l},\tilde{m})$.
Dots below the invariant curve, Eqns.~(\protect\ref{c1}) and
(\protect\ref{c3}), are mapped onto dots above the curve.
\label{horncurve}}
\end{figure}
%
%
\begin{figure}
\centering
\BoxedEPSF{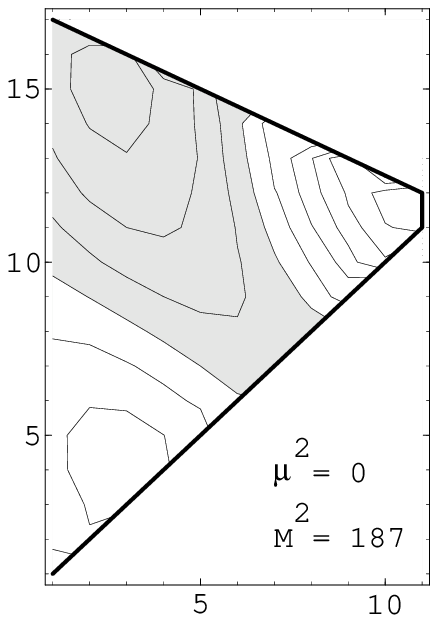 scaled 1200} \BoxedEPSF{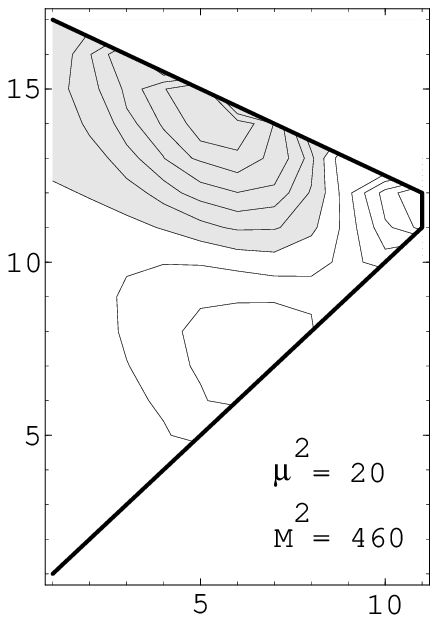 scaled 1200}
\BoxedEPSF{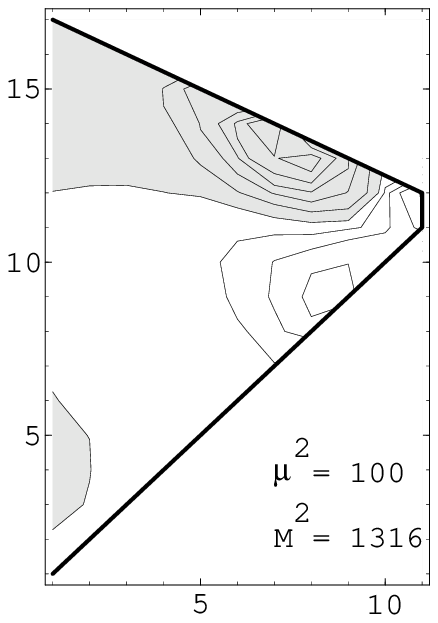 scaled 1200} \BoxedEPSF{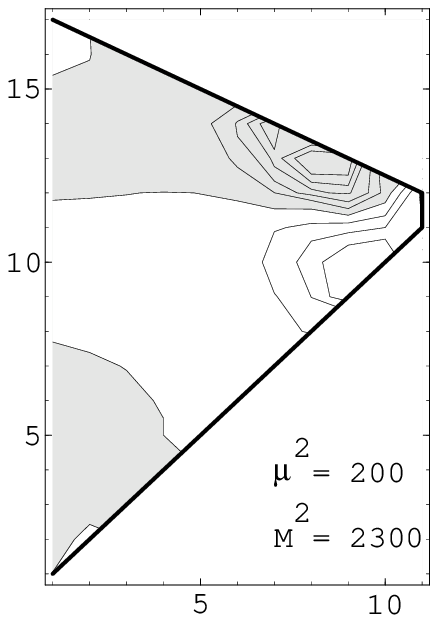 scaled 1200}
\caption{Contour plots of the wavefunction $\phi_{lm}$
vs $l$ and $m$ for the lowest
${{\cal J}_z}^{\cal PC}=3^{+-}$ state and various
values of $\mu^2$.
Regions of negative amplitude are shaded gray.
Comparing this plot with Fig.\ \protect\ref{horncurve},
we see that the wavefunction becomes an approximate
eigenstate of Hornbostel's parity operator as $\mu^2$ is increased.
\label{wfexample}}
\end{figure}
Next, we discuss our method for measuring parity using the
three particle spectrum as an example.
Since the parity operator exchanges $x^+$ and $x^-$, it
is a dynamical operator in light-front quantization and is, in general,
difficult to construct.  In addition, parity is broken
by our DLCQ cutoff and can only be recovered in the continuum limit.
(We apply periodic boundary conditions in the $x^-$
direction, but not in the $x^+$ direction.)
However, in
a theory without fermions we expect that parity will be
restored automatically in the continuum limit.
To determine the parity of a state, we first determine its parity
for finite but large $\mu^2$
using Hornbostel's parity transform for free fields~\cite{hornbostel};
we then decrease the mass to the value of interest.

Three particle eigenstates have the
general form:
\begin{equation}
    \left( f_{abc}\; \mbox{or}\; d_{abc}\right) \sum_{l\le m} \sum_n
    \delta_{K,l+m+n}\,\phi_{l m}\,
    {a_{l,1}^a}^{\dag} {a_{m,1}^b}^{\dag}
    {a_{n,-1}^c}^{\dag} |0\rangle
    \;, \;\;\;\; {\cal J}_z=1 \label{b1}
\end{equation}
and
\begin{equation}
    \left( f_{abc} \;\mbox{or}\; d_{abc}\right) \sum_{l \le m \le n}
    \delta_{K,l+m+n} \, \phi_{l m} \,
    {a_{l,1}^a}^{\dag} {a_{m,1}^b}^{\dag} {a_{n,1}^c}^{\dag} |0\rangle
    \;, \;\;\;\; {\cal J}_z=3 \; . \label{b3}
\end{equation}
In Ref.\ \cite{hornbostel} Hornbostel introduces a parity transform
(times a boost) that is valid for free fields.
In general, a particle with
longitudinal momentum fraction $x_i$ transforms as
\begin{equation}
  x_i \rightarrow \tilde{x}_i =
  	\frac{1}{x_i \sum_j \frac{1}{x_j}} \; .
\end{equation}
In Fig.~\ref{horncurve} we apply this transformation to Eqns.~(\ref{b1}) and
(\ref{b3}).  In each case there is a curve that is invariant under the
transformation:
\begin{equation}
  m = K - \frac{l}{2} - \sqrt{l \left(K-\frac{3}{4} l\right)}
  \; , \;\;\;\; {\cal J}_z=1
  \label{c1}
\end{equation}
and
\begin{equation}
  m = \sqrt{l \left(K-\frac{3}{4} l\right)}-\frac{l}{2}
  \; , \;\;\;\; {\cal J}_z=3 \; .
  \label{c3}
\end{equation}
For free fields, states whose
wave functions $\phi_{l m}$ are even (odd) under this transform have
even (odd) ${\cal PC}$.

Since we are studying an interacting theory, we do not expect
Hornbostel's parity operator to work very well for small $\mu^2$.
This is, in fact, the case.  On the other hand, as we increase $\mu^2$, we
see that the eigenfunctions tend to become approximate eigenstates
of Hornbostel's parity operator.
Thus, we determine the parity of a state by applying Hornbostel's
parity operator to the state for large $\mu^2$.
However, in the large mass limit $\mu^2\to \infty$, finite $K$
effects become important:  eigenstates become states of definite
momenta and the expectation value of Hornbostel's parity
operator vanishes.  Fortunately, there
is a mass region $200 < \mu^2 < 1000$ where Hornbostel's
parity operator does work fairly well.
For example, if we compare Figs.~\ref{horncurve} and \ref{wfexample},
we see that, as $\mu^2$ increases, the state becomes
more approximately an eigenstate of Hornbostel's parity operator.\\

%
\begin{figure}
\centering
\BoxedEPSF{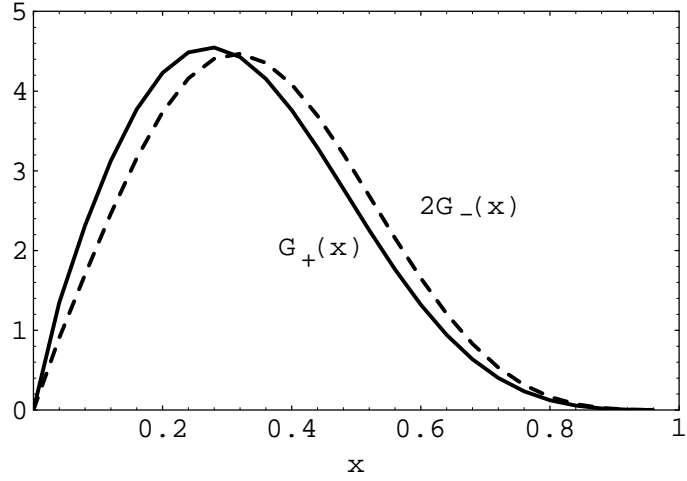 scaled 1200}
\caption{Polarized structure functions of the lowest
$1^{--}$ state for $K=25$ and $\mu^2=0$.
\label{sf1}}
\end{figure}
\begin{figure}
\centering
\BoxedEPSF{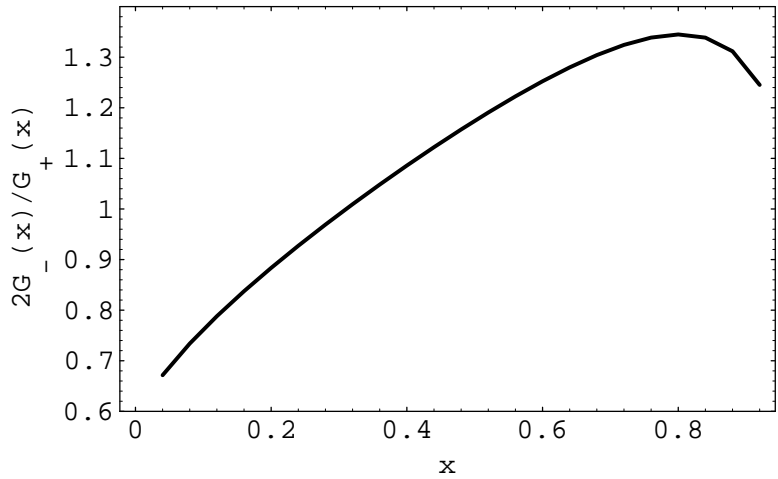 scaled 1200}
\caption{Ratio of the structure functions in
Fig.~\protect\ref{sf1}.  The antiparallel gluon
distribution $G_-(x)$ is somewhat ``harder'' than the
parallel gluon distribution $G_+(x)$. \label{sf2}}
\end{figure}
Finally, we examine the polarized structure functions
for the lowest $1^{--}$ state in Figs.~\ref{sf1} and \ref{sf2}.
We see that the antiparallel gluon has a somewhat harder
momentum distribution than the two parallel gluons.
This is the opposite of what one would normally expect on the
basis of helicity retention \cite{retention}.
Physically, this result is related to the fact that the $2^{++}$ glueball
is lighter than the $0^{++}$ glueball. The reason is the
following: If one thinks of the $1^{--}$ glueball as a
gluon-digluon bound state, then the digluon is lighter when it
is in a $2^{++}$ state than when it is in a $0^{++}$ state.
The ${\cal J}_z=2$ digluon requires an antiparallel third gluon
to yield total ${\cal J}_z=1$, while the ${\cal J}_z=0$
digluon requires a parallel gluon.
Since the ${\cal J}_z=2$ digluon is lighter, it carries
less momentum fraction than the ${\cal J}_z=0$ digluon, and
hence the antiparallel
third gluon tends to be shifted towards larger momenta than
the parallel third gluon.
Of course, this argument oversimplifies the situation
somewhat since there is spin dependence in the interaction
between the ``digluon'' and the ``third'' gluon. Furthermore,
since the gluons are indistinguishable one cannot really separate
them into a ``digluon'' and a ``third gluon''.
Nevertheless, our main point should still be correct:
that there is a connection between the $2^{++}$ glueball
being lighter than the $0^{++}$ glueball
and having antiparallel gluons that are ``harder'' than parallel
gluons.

\section{Four particles}
\label{fourp}

When one includes Fock components with a maximum of four gluons,
several new phenomena occur:  states that were present already
within the two particle truncation mix with states that
have four particles and new states appear
that have little or no two particle content.
Although the number of colors $N$ must
be specified for the three particle truncation, the
four particle sector is the first place where the value
of $N$ can have an important effect on the spectrum.
%
%
\begin{figure}
\centering
\BoxedEPSF{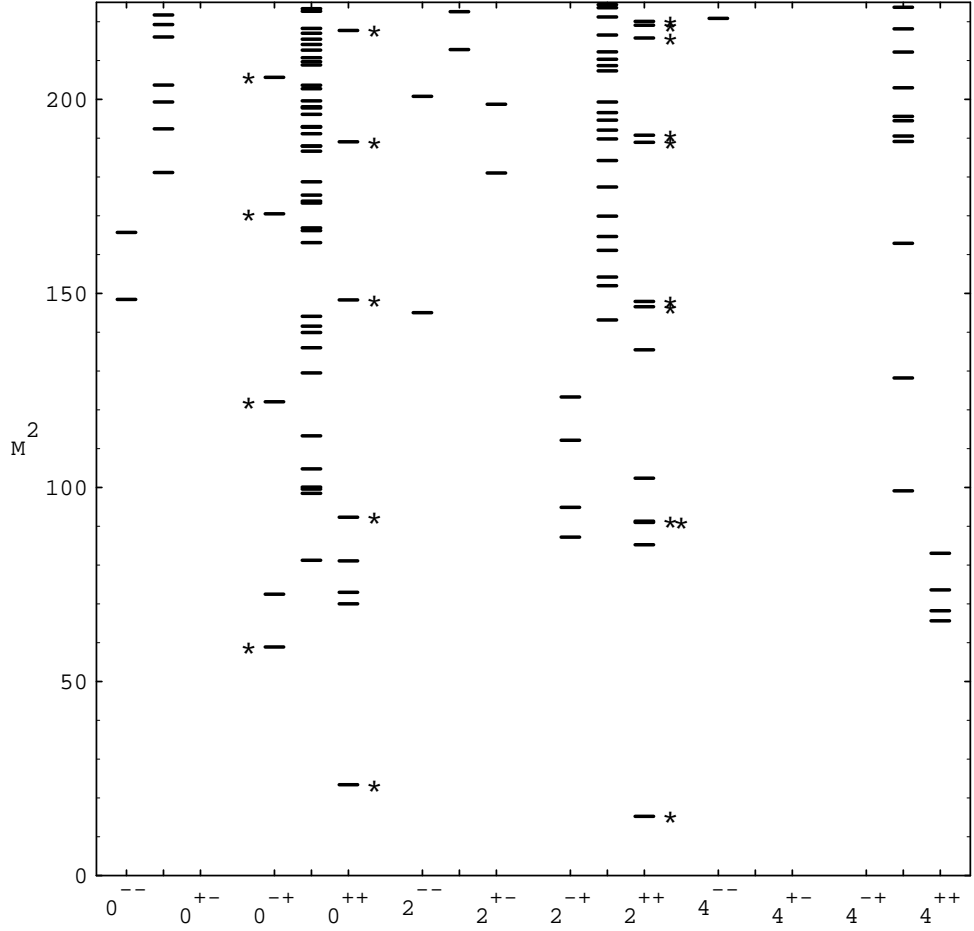 scaled 1000}
\caption{Four particle spectrum for various ${{\cal J}_z}^{PC}$;
$K=12$ (${\cal J}_z=0,\,2$) or $K=17$ (${\cal J}_z=4$), and $\mu^2=0$.
States whose parity could not be determined
are placed midway between  the corresponding ${{\cal J}_z}^{\cal +C}$ and
${{\cal J}_z}^{\cal -C}$ columns.
States with more than 25\% two particle content are marked with an
asterisk.   \label{fig4spec}}
\end{figure}
The spectrum is shown in Fig.~\ref{fig4spec};
values of $M^2$ quoted later in this section are based on this spectrum.

The ${\cal C}=-1$ sector contains states that are rather high
in energy.  These states of course have no two particle content.
It is rather puzzling that the lowest state in each ${\cal C}=-1$
sector has odd parity; this violates the symmetry principle
that we discussed in Section \ref{threep}.

%
\begin{figure}
\centering
\BoxedEPSF{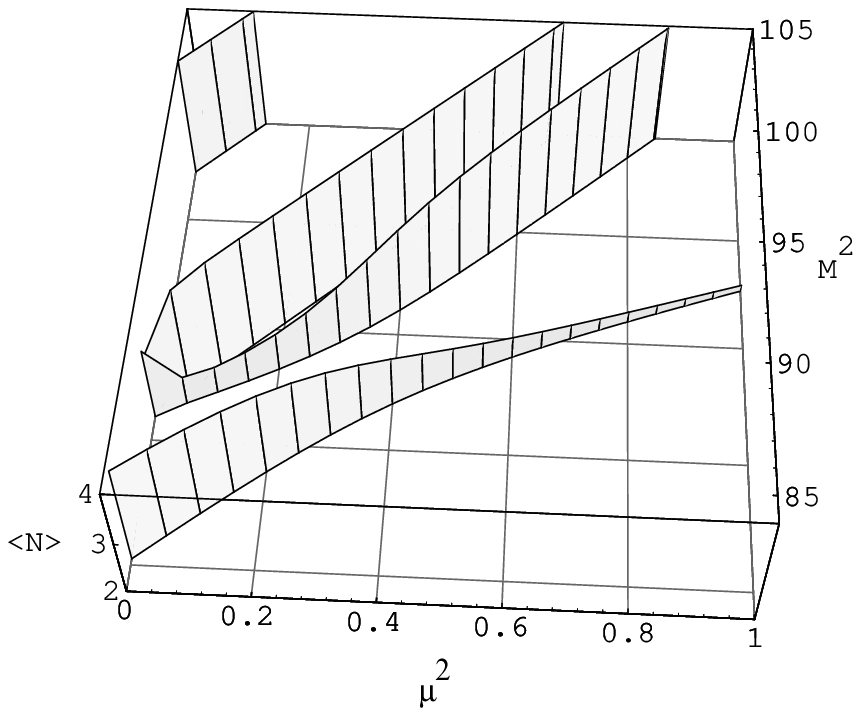 scaled 1200}
\caption{Expectation value of the number operator $\langle N\rangle$ vs
$M^2$ and $\mu^2$ for the $2^{++}$ sector of four particle
spectrum, $K=12$.  We see a classic example of level
repulsion as the two and four particle states cross.
\label{twofour}}
\end{figure}
Let us turn our attention to the states with significant two
particle content (marked by asterisks in Fig.~\ref{fig4spec}).
Surprisingly, adding extra gluons has only very little
effect on the masses and structure functions of the lightest
glueballs (typically only a few percent change).
Similar observations had been made in
$QCD_{1+1}$ \cite{hornbostel,mbnuclphysa}.
The reason for this suppression is
not
completely understood but has to do with the
absence of UV divergences (the model is superrenormalizable)
and IR divergences (which are screened in color singlet
bound states) which normally drive the production of
higher Fock components. In combination with the
energy gap (due to the creation of an additional
${\cal J}_z =0$ glueball) that has to be overcome to mix in higher
Fock components, these effects are very suppressed for
low lying glueballs.  Of course the situation changes
when we examine heavier states.
In that case there can be significant mixing of two and four particle
wavefunctions.

An example of mixing of two and four particle wavefunctions
is illustrated in Fig.~\ref{twofour}.  At $\mu^2=0$, the
$2^{++}$ sector of the theory has two nearly degenerate states
at $M^2=91.0$ and $91.3$.  Both of these states have significant
two particle content.  In addition, there is a four particle state
at $M^2=85.2$. As we increase $\mu^2$ to 1, the two particle content
of the two upper states decreases and the lower state becomes
mainly two particles.  Thus, we see a classic example of
``level repulsion'' as the two particle state crosses the
four particle state.
The strength of the interaction between the two and
four particle states is roughly
$\left(M^2\right)_{2,4} \approx 2$.

%
\begin{figure}
\centering
\BoxedEPSF{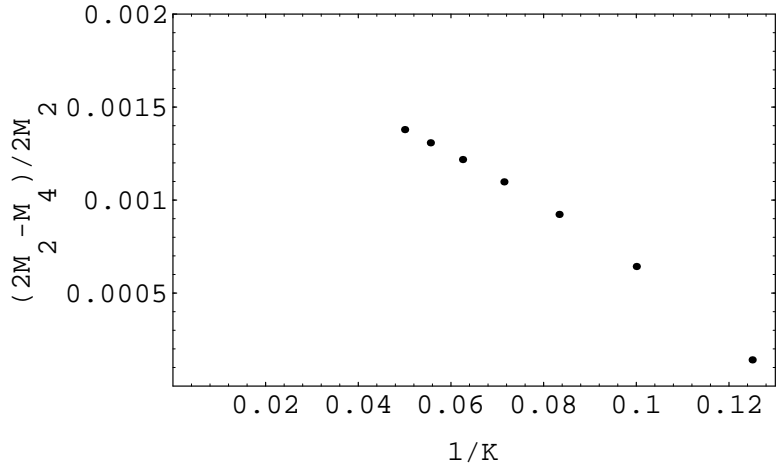 scaled 1200}
\caption{Binding energy of the lightest
$4^{++}$ glueball in units of the mass of
two $2^{++}$ glueballs as a function of the inverse
harmonic resolution for $\mu^2=1$. \label{figa}}
\end{figure}
\begin{figure}
\centering
\BoxedEPSF{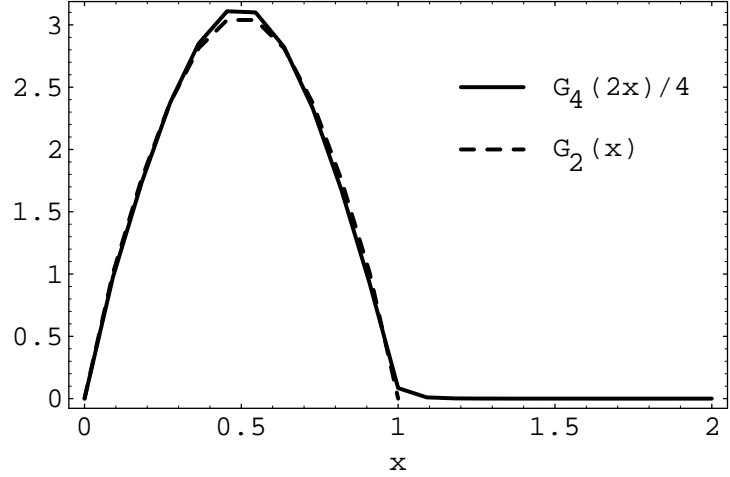 scaled 1200}
\caption{Gluon distributions for the $2^{++}$
glueball (dashed curve, $K=10$, and $\mu^2=1$) and
the $4^{++}$ glueball (solid curve, $K=20$, and $\mu^2=1$).
We have rescaled the $4^{++}$ gluon
distribution $G_4(x)$ such that it would
yield the same distribution as the
$2^{++}$ glueball in the zero binding limit.\label{figb}}
\end{figure}
\begin{figure}
\centering
\BoxedEPSF{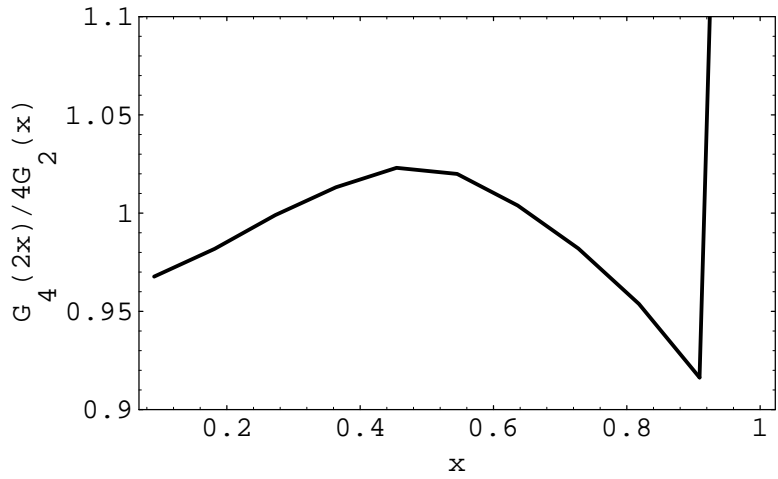 scaled 1200}
\caption{Ratio of the gluon distributions shown in
Fig.~\protect\ref{figb}. \label{figc}}
\end{figure}
Finally, we consider the ${\cal C}=1$ states with predominately
four particle content.  Let us start with the
lightest $4^{++}$ glueball ($M^2=66$) as an example.
Numerical calculations at various
values of the harmonic resolution $K$ (Fig.~\ref{figa})
show that its mass is slightly lower than the mass
of two $2^{++}$ glueballs; that is, decay into two
widely separated $2^{++}$ glueballs is energetically forbidden and
the lowest $4^{++}$ glueball is stable.
In order to minimize numerical artifacts in the
analysis of this state, we compare in the following
the results for the $4^{++}$ glueball taken at $K=20$
with those of the $2^{++}$ glueball at half that value ($K=10$).
Furthermore, since no additional gluons (e.g. a 6 particle Fock
component) were allowed, we compare with the $2^{++}$ glueball
restricted to the two gluon Fock sector.

Since its mass is extremely close to the mass
of two $2^{++}$ glueballs, one would expect that this state
closely resembles a loosely bound pair of $2^{++}$
glueballs.
A comparison of the gluon distributions in Figs.~\ref{figb}
and \ref{figc}
confirms this effect since the distribution of the
$2^{++}$ and the (appropriately rescaled) $4^{++}$
glueball are almost identical. The most pronounced
differences are:
\begin{itemize}

\item a slight enhancement
around $0.4<x<0.5$ which we interpret as a Bose-effect:
when one brings two $2^{++}$ glueballs together,
there is an enhancement of the gluon distribution
in the region of the maximum due to the symmetrization
of the gluon wavefunction.

\item a tail that goes beyond $x \geq 1$ in the case
of the $4^{++}$ glueball, which is kinematically forbidden
for the $2^{++}$ glueball. This behavior is also reflected in the
divergence of the ``EMC-ratio'' at $x=1$ in Fig.~\ref{figc}.

\item  a depletion for $x<0.4$ and for $0.5<x<0.95$:
the momentum sum rule requires that the EMC-ratio drops
below 1 at least once, which would
``explain'' the dip near $x=0.9$. However, we have not
been able to find a similarly simple reason to explain
the drop for $x<0.4$. See, however, the discussion
of the deuteron in Ref.\cite{mbnuclphysa}
which showed a similar behavior.

\end{itemize}
In summary, the $4^{++}$ glueball has much in common
with a ``molecule'' or an ``atomic nucleus'' in the
sense that many of its properties can be explained
by a picture of weakly bound constituents
(here two $2^{++}$ glueballs). Numerical limitations
did not allow us to study whether similar bound states
appear for ${\cal J}_z =6,8,10,...$.
Intuitively one would expect this
to happen since (1) already weak attractive forces lead to binding in
1 spatial dimension and (2) gluons are bosons and
one would not expect a shell structure such as the one
which occurs in $QCD_{1+1}$ \cite{mbnuclphysa}.

We can apply a similar analysis to the ${\cal J}_z=2$ and
${\cal J}_z=0$ sectors of the four particle spectrum.
We find that the lowest $2^{++}$ glueball with mainly
four particle content ($M^2=85$) is a
weakly bound state of a $0^{++}$ and a $2^{++}$ glueball.  Likewise the
lowest $0^{++}$ state with mainly four particle content
($M^2=70$) consists of two
weakly interacting $2^{++}$ glueballs.
However, in this case, the state is not bound.
In fact, we can also identify the lowest $0^{++}$ state consisting
of two weakly interacting $0^{++}$ glueballs at $M^2=105$.
This state is not bound either.

The excited states in the four particle spectrum
consist mainly of pairs of weakly interacting
glueballs which are moving at different relative momenta.  In fact,
with possibly one exception, all of the states below $M^2 <140$
have this interpretation.
For this reason, the levels shown in Fig.~\ref{fig4spec} are
well approximated by the DLCQ formula for two free particles
of invariant mass $M_1$ and $M_2$,
\begin{equation}
  M^2 = K \left(\frac{M_1^2}{i}+\frac{M_2^2}{K-i}\right)
   \; ,
\end{equation}
and therefore the level spacing of these states is a finite $K$ artifact.
The physically correct level spacing would be much smaller for the bound
states and there would be a continuum spectrum in the $0^{++}$ sector.

Because of this ``continuum'' of weakly interacting pairs of glueballs,
it is somewhat difficult to distinguish any ``true'' four particle states.
The only candidate that we could identify
is a $0^{++}$ state at $M^2=98$.  If we were to take the $N \to \infty$
limit, this would be the lowest four particle state that would survive.

\section{Discussion}
\label{discussion}

How well does the tube model compare with experiment?
The most reliable results that we have to compare to
are from lattice gauge theory~\cite{lattice}.  The most striking
feature of our spectrum is the fact that the $2^{++}$ glueball
is lighter than the $0^{++}$ glueball.
This result is surely not correct.
In principle, the coupling associated with the $\left[A_r,A_s\right]^2$
interaction in our Hamiltonian is independent of the coupling
associated with the instantaneous interactions $P^-_1 + \cdots +P_4^-$.
Thus, for instance, one might adjust these
couplings independently to obtain a spectrum that is more physically
realistic \cite{dalley}.  However, we note two problems with this scheme.
First, we point out the special behavior of the
two particle spectrum at $\mu^2=0$ (namely, the constant wavefunction
and zero eigenvalue) occurs only when the two couplings are identical.
Moreover, in order to have better fit to lattice data,
for instance,
we would need to change the {\em sign} of the $\left[A_r,A_s\right]^2$
coupling.
With such a sign change, the spectrum may be no longer be
bounded from below.  This is especially a concern if one attempts to
adjust the couplings by fitting the lowest $0^{++}$ and $2^{++}$
glueballs to the lattice data.

Another concern is the slow convergence of the DLCQ calculations
near $\mu^2 =0$.  The simplest improvement would be to choose
antiperiodic instead of periodic boundary conditions for
the gauge fields.  However, if
we later want to include fermions in the theory, we must choose periodic
boundary conditions for the gauge fields.  Another approach would be
to solve the constraint equation for the zero mode of $A_r$
and substitute the result back into the Hamiltonian.
This would produce new operators in the Hamiltonian and would
substantially improve convergence to the continuum limit.
This technique has already been demonstrated for $\phi^4$ theory
in 1+1 dimensions \cite{phi4}.  Finally, one can simply abandon
DLCQ as a numerical technique and use a polynomial wave
function basis instead; this is what we did for the two particle
truncation.  However, extending such a technique to a many particle
calculation is relatively difficult.

In conclusion, we have made a thorough study of the
3+1 dimensional tube model.
In particular, we have examined the behavior of the $\mu^2 \to 0$
limit and studied the convergence of the DLCQ calculations.
In addition, we have successfully introduced an empirical
technique to measure the parity of a state.
We have found a discrete spectrum in the two and three particle
sectors along with a continuous spectrum in the four particle sector
and have studied the mixing of two and four particle states.

\section*{Acknowledgements}
The authors would like to thank R. Bayer, S. Dalley, A. Kalloniatis,
and H.-C.\ Pauli for useful discussions and comments.
M. Burkardt thanks the MPI f\"ur Kernphysik in Heidelberg
for its hospitality.
This work was supported, in part, by the Alexander Von Humboldt Stiftung.

\end{document}